\documentclass[hidelinks,journal]{IEEEtran}
\usepackage{cite}
\usepackage{amsmath}
\usepackage{array}
\usepackage{color}

\ifCLASSOPTIONcompsoc
  \usepackage[caption=false,font=normalsize,labelfont=sf,textfont=sf]{subfig}
\else
  \usepackage[caption=false,font=footnotesize]{subfig}

\usepackage{dblfloatfix}
\usepackage{url}

\usepackage{color}
\usepackage{amsthm}
\usepackage{amsmath}
\usepackage{amssymb}
\usepackage{amsfonts}
\usepackage{csquotes}
\usepackage{mathrsfs}
\usepackage{mathtools}
\DeclarePairedDelimiter\ceil{\lceil}{\rceil}
\usepackage{graphicx}
\usepackage{ragged2e}
\usepackage{footnote}
\usepackage[ruled,linesnumbered]{algorithm2e}
 \usepackage{algpseudocode}
\usepackage[normalem]{ulem}
\usepackage{dirtytalk}

\usepackage{enumitem}	% For modifity left margins % http://tex.stackexchange.com/questions/170525/itemize-left-margin
\usepackage{verbatim}
\usepackage{booktabs}
\usepackage{float}

\usepackage[T1]{fontenc}
\usepackage{authblk}

\usepackage[pdfpagelabels,hypertexnames=false,breaklinks=true,bookmarksopen=true,bookmarksopenlevel=2]{hyperref}

\renewcommand{\IEEEQED}

\usepackage{hyperref}

\makeatletter
\let\original@algocf@latexcaption\algocf@latexcaption
\long\def\algocf@latexcaption#1[#2]{%
	\@ifundefined{NR@gettitle}{%
		\def\@currentlabelname{#2}%
	}{%
		\NR@gettitle{#2}%
	}%
	\original@algocf@latexcaption{#1}[{#2}]%
}
\makeatother

\begin{document}
\title{An Asynchronous Multi-Beam MAC Protocol for Multi-Hop Wireless Networks}

\author{Shivam~Garg\thanks{S. Garg is with the Computational Science Research Center (CSRC) at San Diego State University (SDSU), CA, USA, sgarg@sdsu.edu},~\IEEEmembership{Student~Member,~IEEE}, \and Nandini~Venkatraman\thanks{N. Venkatraman is with the Electrical and Computer Engineering (ECE) Department, SDSU, San Diego, CA, USA, nanvenkatram@gmail.com}, \and Elizabeth~Serena~Bentley\thanks{E. S. Bentley is with the Air Force Research Laboratory, Rome, NY, USA, elizabeth.bentley.3@us.af.mil},~\IEEEmembership{Member,~IEEE}, \and Sunil~Kumar\thanks{S. Kumar is with the CSRC and ECE Department, SDSU, San Diego, CA, USA, skumar@sdsu.edu

DISTRIBUTION STATEMENT A: Approved for Public Release; distribution unlimited AFRL-2021-3714 on 25 Oct 2021. Other requests shall be referred to AFRL/RIT 525 Brooks Rd Rome, NY 13441
},~\IEEEmembership{Senior~Member,~IEEE} }

\maketitle

\begin{abstract}

A node equipped with a multi-beam antenna  can achieve a throughput of up to $m$ times as compared to a single-beam antenna, by simultaneously communicating on its $m$ non-interfering beams. However, the existing multi-beam medium access control (MAC) schemes can achieve concurrent data communication only when the transmitter nodes are locally synchronized. Asynchronous packet arrival at a multi-beam receiver node would increase the node deafness and MAC-layer capture problems, and thereby limit the data throughput. This paper presents an asynchronous multi-beam MAC protocol for multi-hop wireless networks, which makes the following enhancements to the existing multi-beam MAC schemes (i) A windowing mechanism to achieve concurrent communication when the packet arrival is asynchronous, (ii) A smart packet processing mechanism which reduces the node deafness, hidden terminals and MAC-layer capture problems, and (iii) A channel access mechanism which decreases resource wastage and node starvation. Our proposed protocol also works in heterogeneous networks that deploy the nodes equipped with single-beam as well as multi-beam antennas. Simulation results demonstrate a superior performance of our proposed protocol.
\end{abstract}


\IEEEpeerreviewmaketitle

%\IEEEpubidadjcol

\section{Introduction}
Directional communication improves the spatial reuse, extends signal transmission range, and reduces interference, thereby increasing the network capacity \cite{Ref1,Ref2}. Therefore, directional communication is used in many different systems, including in 5G systems, radar systems, wireless LANs, mobile ad hoc and airborne networks \cite{Ref3,Ref4,Ref5,Ref6,Ref7,Ref8,Ref9,Ref10}. At the same time, recent advances in antenna technology, along with a shift toward higher frequencies, have made the multi-beam directional antennas (MBA) more feasible. As a result, the use of wireless nodes equipped with MBA has attracted considerable interest in the literature \cite{Ref2,Ref3,Ref4,Ref5,Ref6,Ref7,Ref9,Ref10,Ref11,Ref12,Ref13,Ref14,Ref15,Ref16}. Since the use of directional antennas divides the space around them in different beams, a multi-beam antenna with \textit{m} beams can allow multiple (up to \textit{m}) simultaneous transmissions or receptions by a node in a single frequency band, thus improving the throughput by up to \textit{m} times as compared to a single beam directional antenna \cite{Ref1,Ref2,Ref11}. 

The network performance can be increased significantly if the routing protocol could optimally use the multiple beams of nodes, which requires the support of a medium access control (MAC) protocol that can enable concurrent communication on the multiple beams of a node \cite{Ref19}. However, the design of  a multi-beam MAC protocol must address the following issues \cite{Ref1,Ref2,Ref10,Ref12,Ref13,Ref16}:

\begin{enumerate}[label=(\roman*)]
 \item A half-duplex multi-beam node can either transmit or receive in multiple beams concurrently, but not both. To enable \textit{concurrent packet reception} (\textit{CPR}), all the transmitting nodes (beamformed in the direction of a given multi-beam receiver node) should start their \textit{transmissions concurrently} (\textit{CPT}) so that the common receiver can simultaneously activate its multiple beams pointing toward them. In other words, the transmitting nodes must be \textit{locally synchronized}.
 
 \item The use of directional communication exacerbates the node \textit{deafness} problem when the intended receiver is beam-formed in a direction away from the transmitter(s). This node deafness decreases the network capacity.
 
 \item The use of directional communication leads to \textit{hidden terminals}, which occurs when a potential interferer does not receive the handshake signals due to its antenna orientation but initiates a transmission that causes the collisions. 
 
 \item A packet at the head of the data queue can block other packets behind it indefinitely if its intended outbound beam is busy. This is known as the \textit{head-of-line} (\textit{HOL}) \textit{blocking} and can be avoided by having different data queues for each beam. 
 
 \item The time a directional node wastes in receiving packets, not intended to it, might refrain it from communicating in other desired directions. This problem is known as \textit{MAC-layer capture} and decreases the spatial reuse.  
 
\end{enumerate}

Existing multi-beam MAC protocols address the above issues only to some extent. In this paper, we describe a novel asynchronous multi-beam MAC scheme for multi-hop wireless networks, which effectively addresses the above issues and achieves a superior performance for both static and mobile network topologies.

\subsection{\textbf{Related Work}}
\label{RelatedWork}
The directional antenna technology can be divided into three broad categories: switched multi-beam antennas, adaptive array antennas, and multiple-input-multiple-output (MIMO) links \cite{Ref1,Ref2,Ref12}. These technologies have been used in various applications, including the radar systems, wireless LANs, mobile ad hoc communication systems, and airborne networks \cite{Ref3,Ref4,Ref5,Ref6,Ref7,Ref8,Ref9,Ref10}.

A review of directional MAC protocols can be found in \cite{Ref1,Ref2,Ref10,Ref12,Ref13,Ref16}, including the important issues that a multi-beam MAC protocol for the wireless networks must address. Note that only a limited number of multi-beam directional MAC protocols are available in the literature \cite{Ref6,Ref7,Ref9,Ref10,Ref11,Ref12,Ref13,Ref14,Ref15,Ref16}; all of which have underlined the importance of CPT and CPR. Some multi-beam MAC schemes use a time division multiple access (TDMA) like synchronous MAC protocol (e.g., \cite{Ref6,Ref7,Ref10,Ref11,Ref14,Ref15}), whereas others use carrier sense multiple access (CSMA) based MAC protocols (e.g., \cite{Ref9,Ref12,Ref13,Ref16}). However, the TDMA based schemes introduce significant overhead and delay in computing a conflict-free transmission schedule in a multi-hop topology. Further, these schemes are not suitable for dynamic networks (such as an airborne network) because the current schedule would become outdated, and a new conflict-free schedule is required whenever the topology changes \cite{Ref20}. 

We consider only CSMA based multi-beam MAC schemes in this paper. A MAC protocol for aerial sensor networks is discussed in \cite{Ref9}, where a main actor unmanned aerial vehicle (UAV) is equipped with an MBA while other UAVs have a single-beam directional antenna. However, \cite{Ref9} is limited to a fixed network topology, and does not address the problems of deafness and hidden terminals. Schemes in \cite{Ref10,Ref12,Ref13,Ref16} support both multi-beam packet transmission (i.e., CPT) and reception (i.e., CPR) for multi-hop wireless networks, and use a channel contention mechanism when multiple nodes in a beam start their transmissions concurrently, which improves fairness. However, schemes in \cite{Ref10,Ref12,Ref13} use both the directional and omnidirectional antennas (for channel sensing and/or access) to avoid the deafness and assume the collision free transmission and reception, which introduces the well-known gain asymmetry problem \cite{Ref1,Ref2}. Whereas the scheme in \cite{Ref16} does not address the issue of MAC-layer capture, which can significantly increase node starvation in a multihop topology and reduce the throughput. 

To the best of our knowledge, none of the existing MAC schemes \cite{Ref6,Ref7,Ref9,Ref10,Ref11,Ref12,Ref13,Ref14,Ref15,Ref16} addresses all the important issues (discussed above) for a multi-beam directional communication. Furthermore, these schemes are not suitable for a mobile and/or random topology, as discussed below. 

\subsection{\textbf{Shortcomings of Existing Multi-Beam MAC Schemes}}
\label{Shortcomings}

\begin{itemize}
\item Most of the multi-beam MAC schemes in literature assume that the transmitter nodes are locally synchronized, and the data packets have the same size so that a multi-beam receiver node can simultaneously receive their transmissions and hence perform CPR. 

\item Furthermore, the propagation delay is ignored in most of the existing schemes. Since the directional antennas facilitate long distance communication links, different propagation delays of randomly positioned mobile nodes can result in asynchronous packet arrival at the multi-beam receiver node. This would lead to node deafness, starvation, and packet collisions, and thereby degrade the CPR capability and throughput, especially in a wide area multihop network. 

\item Use of different packet sizes by different flows in a network can also result in asynchronous packet arrival at a multi-beam receiver node due to different transmission delays, which would also limit CPR and throughput.
\end{itemize}

%If a node does not achieve successful transmission in a few consecutive attempts, it can mistakenly flag ‘network congestion’ to request upper layer(s) to start a fresh route discovery and/or adjust the flow control.

\subsection{\textbf{Contributions of This Paper}}
\label{Contributions}
Our proposed asynchronous multi-beam MAC protocol completely resolves all the above-mentioned issues for multi-hop wireless networks which consist of a mobile and random network topology. It has the following important features.

\begin{enumerate}[label=(\roman*)]
\item A windowing mechanism is introduced to achieve the CPT and CPR when the packet arrival is asynchronous due to random and mobile topologies, and different data packet sizes. 

\item A smart packet processing mechanism is introduced which reduces the node deafness, hidden terminals and MAC-layer capture problems. As a result, the need for packet retransmission and resulting overhead are reduced. 

\item A new channel access mechanism is introduced which decreases the resource wastage and node starvation. 

\item Our proposed scheme also works in the networks that deploy the nodes equipped with single-beam and multi-beam antennas.
\end{enumerate}

This paper is organized in five sections. An overview of existing multi-beam MAC protocols is discussed in Section \ref{OverviewBasicMultiBeamSchemes}. Our proposed multi-beam MAC protocol is discussed in Section \ref{OurMultiBeamScheme}, including a better resource management approach, windowing algorithm, and an improved channel access mechanism, followed by its working principle. The simulation results and performance are discussed in Section \ref{SimualtionResults}, followed by the conclusions in Section \ref{Conclusion}.

\section{OVERVIEW OF BASIC MULTI-BEAM MAC PROTOCOL}
\label{OverviewBasicMultiBeamSchemes}
The IEEE 802.11 DCF (distributed coordination function) protocol was designed for omnidirectional communication. A multi-beam node can transmit or receive multiple packets concurrently on its different beams. However, all beams of a half-duplex multi-beam node can either be in transmission (CPT) or in reception (CPR) mode at a given time \cite{Ref9,Ref10,Ref11,Ref16}. The MBA is assumed to be an ideal wide-azimuth antenna which covers the ${360^0}$ through its \textit{m} non-overlapping beams such that each beam covers $\frac{360^0}{\textit{m}}$ angle. The total transmit power is equally divided among these \textit{m} beams, and independent channel sensing is performed for each beam. The key features of the existing multi-beam MAC protocols are discussed below \cite{Ref6,Ref7,Ref9,Ref10,Ref11,Ref12,Ref13,Ref14,Ref15,Ref16}. 

To achieve CPT, all the active beams of a multi-beam node must be synchronized \cite{Ref9,Ref10,Ref11,Ref16}. To achieve CPR, a multi-beam receiver node also requires all the intended transmitter nodes to transmit simultaneously towards its different beams. Since the node deafness and hidden terminal problems increase significantly in directional communication, a multi-beam node should transmit control packets on all those beams which contain a potential transmitter for it \cite{Ref9,Ref10,Ref16}. The IEEE 802.11 DCF MAC uses a 4-way handshake (RTS, CTS, Data, and ACK) to minimize the collision probability, where the contention window (CW) of a node decreases to the minimum after each successful transmission, whereas it is doubled for each unsuccessful transmission until the maximum retry limit is reached. During concurrent communication on its \textit{m} beams, a multi-beam node may experience some unsuccessful transmissions. To address this issue, the node-based backoff is used, in which a multi-beam node maintains a single CW for its all beams \cite{Ref9,Ref16}. If at least one beam achieves successful transmission, the multi-beam node’s CW is set to minimum. To avoid the HOL packet blocking problem \cite{Ref1,Ref2}, a multi-beam node maintains a separate data buffer queue for each beam at the MAC layer so that it can concurrently transmit packets on its multiple beams.

The following three changes are made to convert the IEEE 802.11 DCF MAC scheme to a basic multi-beam MAC scheme. 

\begin{enumerate}[label=(\roman*)]
\item \textit{\textbf{Removal of Random Backoff}} - All transmitter nodes should transmit simultaneously to achieve CPR at the multi-beam receiver node. Therefore, random backoff is removed so that the transmitter nodes can be locally synchronized \cite{Ref9,Ref10,Ref16}.

\item \textit{\textbf{Identification and Handling of Potential Transmitter Nodes}} - Each multi-beam node maintains a modified NAV (network allocation vector) table which contains the per-beam information \cite{Ref9,Ref10,Ref16}. For example, this table in \cite{Ref16} contains the neighbor node’s address, NAV, and flags for potential transmitters. If a multi-beam node receives an RTS (or CTS) from a potential transmitter (or receiver) node on its beam, it sends a scheduling RTS or CTS packet and updates the NAV of that beam (for beam synchronization purpose). Upon receiving such a packet, the neighbor node(s) updates its NAV for the duration until it needs to defer its transmission. 

\item \textit{\textbf{Channel Access Strategy}} - Without the random backoff, all locally synchronized nodes start their transmissions concurrently \cite{Ref10,Ref13,Ref16}. To allow a multi-beam receiver node to access the channel, a role priority switching is used, where a node performs successive cycles of CPT and CPR. It enables a receiver node to start its own transmission.  

\end{enumerate}

\begin{figure}
	\centering
	\includegraphics[width=\linewidth]{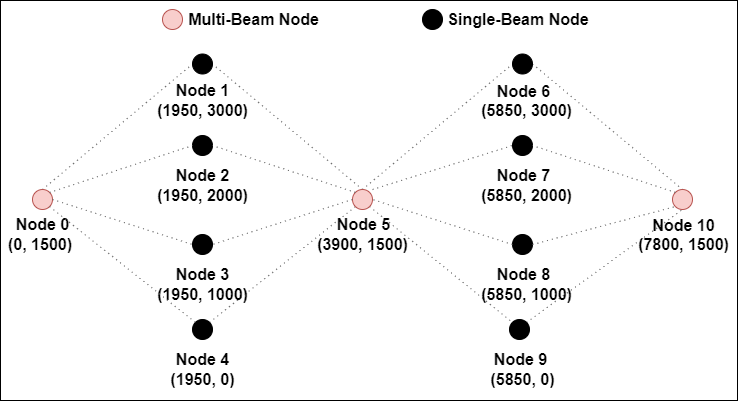}
	\caption {Nodes in a 4-hop topology, where nodes 0, 5, and 10 use a multi-beam antenna. Here (x, y) represent the node location in meters. }
	\label{Fig1}
\end{figure}

\section{OUR PROPOSED MULTI-BEAM MAC PROTOCOL}
\label{OurMultiBeamScheme}
Our proposed multi-beam MAC protocol is designed for random, mobile and multihop wireless network topologies. It uses smart packet processing, which reduces resource wastage and improves data throughput (see Section \ref{SmartPacketProcessing}). A windowing mechanism is used (see Section \ref{AsynchronousPacketArrival}) that allows multi-beam nodes to concurrently receive packets that may arrive asynchronously on their different beams due to different packet sizes and propagation delays. 
%Moreover, it can be used for both shorter as well as longer transmission ranges because of the flexibility provided by the windowing mechanism (discussed in Section \ref{AsynchronousPacketArrival})
Further, an improved channel access mechanism is used (discussed in Section \ref{ImprovedChannelAccess}), which decreases the node deafness and starvation. To reduce the hidden terminal nodes, our proposed scheme uses a mechanism (like in \cite{Ref16}), wherein a multi-beam node transmits the notification RTS and CTS packets (which we call N-RTS and N-CTS, respectively) to inform its potential transmitter node(s) about its ongoing communication. We assume that each node knows its 1-hop neighbors through periodic neighbor discovery.

\subsection{Smart Packet Processing to Improve Resource Management}
\label{SmartPacketProcessing}
In this section, we discuss the inability of the existing multi-beam MAC protocols to handle the arrival of undesired\footnote{The classification of desired and undesired packet is discussed in the $1^{st}$ paragraph of Section \ref{WindowingAlgo}.}  packets, followed by our proposed packet processing mechanism. Recall that the existing multi-beam MAC schemes cannot handle asynchronous packet arrival. In Fig. \ref{Fig1}, consider a situation where multi-beam node 5 transmits its packets to nodes 2 and 3, and has not received an RTS packet on its beam directed towards node 6. As a result, node 5 does not know that node 6 could be a potential transmitter to it. Therefore, it does not send an N-RTS packet towards node 6 while communicating with nodes 2 and 3. If node 6 has a packet for node 5, it can result in one of the following two cases: 

\begin{enumerate}[label=(\roman*)]
\item If node 5 receives CTS packets concurrently from nodes 2 and 3, followed by an RTS packet from node 6 after some duration (< SIFS (short inter frame space)), it discards the RTS packet of node 6. As a result, node 6 doubles its CW and retransmits RTS to node 5. This can eventually lead to the maximum retry limit, followed by a request to the upper layer(s) for a fresh route discovery.  

\item If the RTS packet from node 6 arrives at node 5 earlier than the CTS packets from nodes 2 and 3, node 5 discards the CTS packets, and mistakenly assumes a packet collision on its beams towards nodes 2 and 3. As a result, node 5 doubles its CW and completes its communication with node 6. Then it retransmits the RTS to nodes 2 and 3. This would reduce the flow throughput from node 5 to nodes 2 and 3.

\end{enumerate}

To address the issue in (i), the multi-beam node 5 in our scheme checks the frame type and intended destination address of the packet(s), which arrive after the CTS packets from nodes 2 and 3 in the above example, but before the SIFS timeout interrupt. Node 5 then updates the potential transmitter information on the receiving beam and sends a notification packet in its next transmission cycle to inform such transmitter node(s) (node 6 in the above example) about its current busy state. After receiving the notification packet, the transmitter node(s) knows when to retransmit and, hence, the resource wastage is avoided. 

To resolve the issue in (ii), our proposed scheme compares the beam index (on which the packet arrives) with those beam indexes on which the node is expecting a response. In the above example, node 5 does not allow the RTS packet from node 6 to cancel the frame timeout interrupt and, therefore, achieves CPR. The issue of MAC-layer capture at nodes 5, 2 and 3 is thus resolved. 

\subsection{Handling the Asynchronous Packet Arrival}
\label{AsynchronousPacketArrival}
The CPR on a multi-beam receiver node in the existing multi-beam MAC schemes (such as \cite{Ref6,Ref7,Ref9,Ref10,Ref11,Ref12,Ref13,Ref14,Ref15,Ref16}) assumes that (\textit{i}) the transmitter nodes are locally synchronized and transmit the packets (where each packet has the same size) at the same time (i.e., CPT) and (\textit{ii}) the distance between every transmitter and receiver node pair is almost the same. However, the use of different packet sizes and/or different propagation delays, due to the non-uniform distances between the transmitter and receiver node pairs and/or multipath propagation, can make the packet synchronization at the receiver (CPR) difficult \cite{Ref10}. As shown in Section \ref{SimualtionResults}, the performance of the existing multi-beam MAC protocol degrades significantly when the packets arrive asynchronously. A windowing algorithm is described below to address this issue.

\begin{figure}[b]
	\centering
	\includegraphics[width=\linewidth]{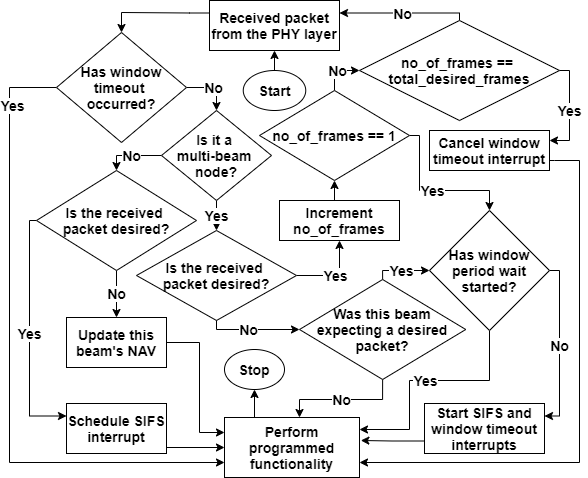}
	\caption {The windowing algorithm to address asynchronous packet arrival.}
	\label{Fig2}
\end{figure}

\subsubsection{Windowing Algorithm}
\label{WindowingAlgo}
The working of the proposed windowing algorithm is shown in Fig. \ref{Fig2}. Here, the packets from the PHY layer are divided into the desired and undesired packets at the MAC layer. A packet is treated as undesired only when: (\textit{i}) A node receives a packet which is not destined to it, or (\textit{ii}) A node transmits RTS (or Data) packets and is waiting for CTS (or ACK) packet(s), but it receives some other packet instead; Otherwise, the packets are treated as desired.

Upon receiving a desired packet, a single-beam node starts its SIFS timer. If an undesired packet is received, the node updates its NAV and performs the programmed functionality. Similarly, a multi-beam node updates its corresponding beam’s NAV on which the undesired packet is received. When a multi-beam node receives its first desired packet, it schedules the SIFS and window timeout interrupts, and then considers only those packets which arrive within the window period duration. The use of a window timeout interrupt ensures a finite window period and prevents the node from suspending its functionality for an indefinite time. Calculation of the window period is discussed in the next section. 

Each multi-beam node maintains the \textit{total\_desired\_frames} and \textit{no\_of\_frames} variables. Here, the \textit{$total\_desired\_frames$} represents the number of packets (except the N-RTS and N-CTS packets) transmitted in the last transmission mode of the current 4-way handshake process, and the \textit{no\_of\_frames} represents the total desired packets received in the current reception mode. For a multi-beam receiver node, which is waiting for an RTS packet, the \textit{total\_desired\_frames} is equal to its multi-beam capacity. The maximum number of simultaneous handshakes in which a multi-beam node can be involved is called its multi-beam capacity. When the \textit{no\_of\_frames} is equal to \textit{total\_desired\_frames}, the multi-beam node cancels its window timeout interrupt.

\begin{figure*}[b]
\begin{equation*}
\tag{2}
Required~role\_switch\_slots = \ceil*{\frac{max(packet~delay~between~the~source~and~intermediate~nodes)}{slot\_time}}
\label{Eq2}
\end{equation*}
\end{figure*}

\begin{figure*}[b]
	\centering
	\includegraphics[width=0.99\textwidth]{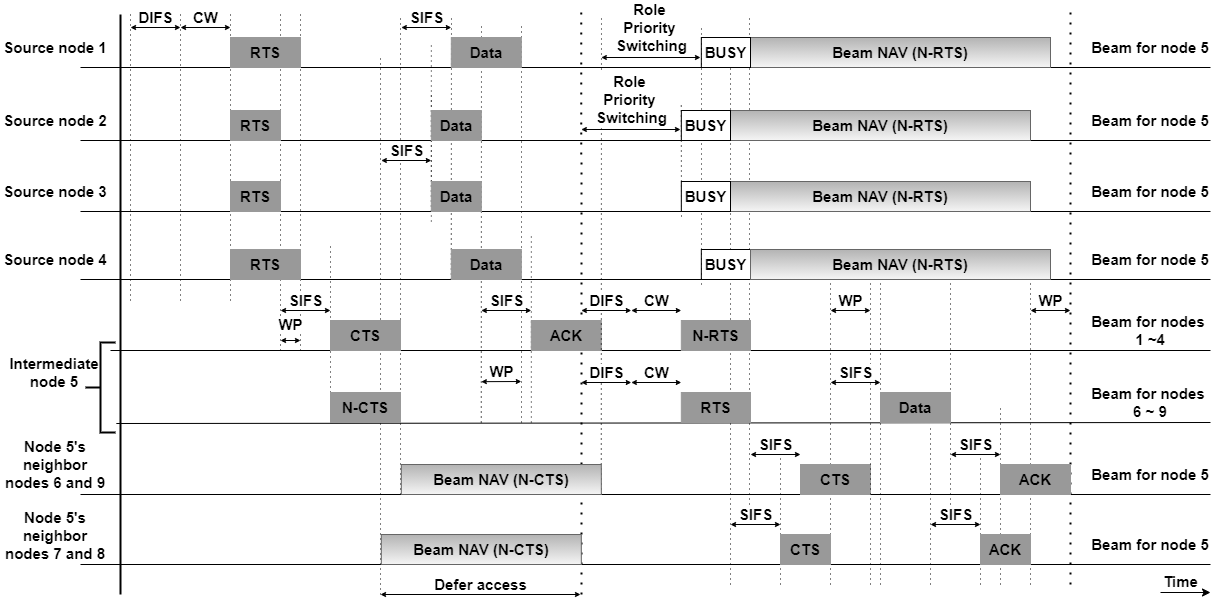}
	\caption {The multi-beam MAC operation in our proposed scheme. Here, the control (RTS, CTS) and data packets are shown to arrive asynchronously at different times.}
	\label{Fig3}
\end{figure*}

\subsubsection{Window Period Calculation}
\label{WindowingPeriodCal}
Upon receiving its first desired packet, a multi-beam receiver node starts its SIFS and window period timers, simultaneously. Because a node waits for the SIFS duration before responding to the received packet(s), the window period should not exceed the SIFS duration as it will increase the packet delay. Since the packet processing incurs non-negligible delay, the window period must be less than the SIFS period, otherwise a multi-beam node may trigger its SIFS interrupt while it is still processing a received packet. We use a window period which is a function of the maximum propagation delay difference between any two locally synchronized nodes \textit{i} and \textit{j}, which have a common multi-beam receiver node, \textit{r}. The window period must satisfy the following condition:
\begin{equation}
\label{Eq1}
2*max(abs(t_{i,r}-t_{j,r})) \leq window~period < SIFS
\end{equation}

Here, $t_{i,r}$ and $t_{j,r}$ represent the propagation delay in covering the distance between nodes \textit{i} and \textit{r} and \textit{j} and \textit{r}, respectively. In Fig. \ref{Fig1}, the nodes 1 through 4 need to be locally synchronized if they want to simultaneously communicate with node 5. Here, the distance of nodes 1 and 4 to node 5 is larger than that of nodes 2 and 3 to node 5. In a 4-way handshake, the RTS packet from node 1 needs abs($t_{1,5}$ – $t_{2,5}$) extra propagation delay (denoted as $t_{propagation\_diff}$) to reach node 5. If node 5 receives all 4 RTS packets within its window period, it can transmit 4 CTS concurrently. Similarly, the CTS packets for nodes 1 and 4 take $t_{propagation\_DIFF}$ extra delay as compared to nodes 2 and 3. The data packet from node 1 or 4 also takes $t_{propagation\_diff}$ extra delay. Therefore, the window period at node 5 must be at least two times of $t_{propagation\_diff}$. 

\subsection{Improved Channel Access Mechanism}
\label{ImprovedChannelAccess}
Due to the removal of the random backoff as discussed in Section \ref{OverviewBasicMultiBeamSchemes}, the existing multi-beam MAC schemes prevent an intermediate node from starvation, by allowing it to alternate between the transmitter and receiver modes \cite{Ref10,Ref12,Ref13,Ref16}. However, a node still prioritizes the transmission mode if it has a packet to transmit in its queue(s), which significantly increases the deafness and node starvation, especially in a multi-hop scenario. 

To address this issue, we use a simple approach, where each node waits for a specific number of time slots (called $role\_switch\_slots$) after successfully receiving ACK. As a result, a node is deferred from accessing the channel for the role switching time. To prevent node deafness and starvation, the intermediate node should access the channel before the source node in the next cycle. Therefore, the value of the $role\_switch\_slots$ depends on the packet delay between the source and the intermediate nodes as shown in Eq. \eqref{Eq2}. 

\subsection{Working of Our Proposed Multi-Beam MAC Protocol}
\label{WorkingOurScheme}
Consider the network topology of Fig. \ref{Fig1}. Here, source nodes 1 through 4 have packets for nodes 6 through 9, respectively. The intermediate multi-beam node 5 has a \textit{multi-beam capacity} of four packets on its four beams and has a potential transmitter on all eight beams. The distance of nodes 1, 4, 6, and 9 from node 5 is 2.5 km, whereas nodes 2, 3, 7, and 8 are 2 km away from node 5. Because of the unequal distances, the use of smart packet processing, windowing, and improved channel access mechanisms are helpful in achieving the concurrent packet transmission and reception. Fig. \ref{Fig3} shows the operation of the proposed multi-beam MAC scheme, which is explained below.

Nodes 1 through 4 sense the channel for the DIFS duration and wait for $CW_{min}$ slots (for constant backoff). Since nodes 1 and 4 are the farthest away from node 5, their packets take a longer time to reach node 5. When node 5 receives its first desired packet, it starts the SIFS and window timeout interrupts. During the window period (WP), node 5 receives all four packets from nodes 1, 2, 3 and 4. Hence, it cancels the window timeout interrupt and waits until the SIFS wait is over. While sending CTS, it sends N-CTS on its remaining beams. Nodes 6 through 9 thus update their NAV and do not interfere with the current communication. Upon receiving the CTS packets, nodes 1 through 4 send Data packets to node 5. Note that the WP time used for the Data packets is twice of the WP time taken by RTS packets. When nodes 1 through 4 receive ACK packets from node 5, they perform the role priority switching. Node 5 then acquires the channel and transmits RTS packets to nodes 6 through 9 and N-RTS to nodes 1 through 4. Nodes 1 through 4 update their NAV and do not interfere with communication between node 5 and nodes 6 though 9.

\section{Simulation Setup and Results}
\label{SimualtionResults}
We have implemented the basic and our proposed multi-beam MAC schemes on the Wireless Suite of Riverbed Modeler in Version 18.0.2 \cite {Ref21}, and their performances are compared for static and mobile network topologies. The implementation details of multi-beam MAC schemes, including the multi-beam antenna module, node model with multiple sources to generate data packets for different destination nodes, and process model to enable CPT and CPR are discussed in \cite{Ref22}. A heterogeneous network is considered, which consists of both the single-beam and multi-beam nodes, where the coverage range of each node is 3 km. A line-of-sight communication is assumed and the channel fading and noise are ignored. The packet size is 1500 Bytes and the simulation is run for 180 s. The beam steering delay is assumed to be negligible. The SRL (short retry limit) for control packets and LRL (long retry limit) for data packets are 7 and 4, respectively. The slot size, SIFS and DIFS duration are 20 µs, 10 µs and 50 µs, respectively. The size of RTS, CTS and ACK packets are 20 Bytes, 14 Bytes and 14 Bytes, respectively. The minimum and maximum values of CW are 16 and 1024, respectively. 

\textbf{Performance Metrics:} The following four performance metrics are used in our simulation:

\begin{enumerate}[label=(\roman*)]
\item Throughput for a flow is the total packets (or bits) received per second by the destination node.

\item PDR (packet delivery ratio) for a flow is the ratio of total packets received by the destination node to the total packets generated at the source node.

\item End-to-end delay is the average time taken by all the data packets to travel from the source node to destination node.

\item Percentage times when extra routes are utilized in our proposed scheme as compared to the basic multi-beam MAC scheme.

\end{enumerate}

\subsection{Performance of Concurrent Packet Communication for Static Network Topology}
\label{StaticScenario}
The CPT and CPR performance of a multi-beam node is studied when the packets arrive at a receiver node asynchronously because the distance between the transmitter and receiver nodes is not uniform. We consider a network topology shown in Fig. \ref{Fig1}, where nodes 0, 5 and 10 are multi-beam nodes and the remaining nodes use a single-beam antenna. The multi-beam node 5 has eight beams, pointing towards nodes 1 to 4 and 6 to 9. The multi-beam nodes 0 and 10 use only four beams which are pointed towards nodes 1 to 4 and 6 to 9, respectively. Here, the performance is compared in terms of throughput when the channel capacity is 5 Mbps. 

Each source generates data at 3 Mbps (250 packets per second @1500 Bytes per packet) on each beam with an inter-arrival time of 4 ms. Since the RTS packet size is 20 Bytes, its transmission delay on a 5 Mbps channel is 32 µs. In addition to it, the maximum propagation delay between nodes 0 and 1 is 8.3 µs in Fig. \ref{Fig1}. Since the total packet delay is 40.3 µs, the multi-beam node must wait for 3 time slots (20 µs per slot) to avoid deafness and/or starvation (see Eq. \eqref{Eq2}). As a result, the role priority switching mechanism uses 3 time slots in our proposed scheme. A window period of 9 µs is used in our simulation, which is lower than the SIFS duration.

\begin{figure}[b]
    \centering
    \includegraphics[width=0.98\linewidth]{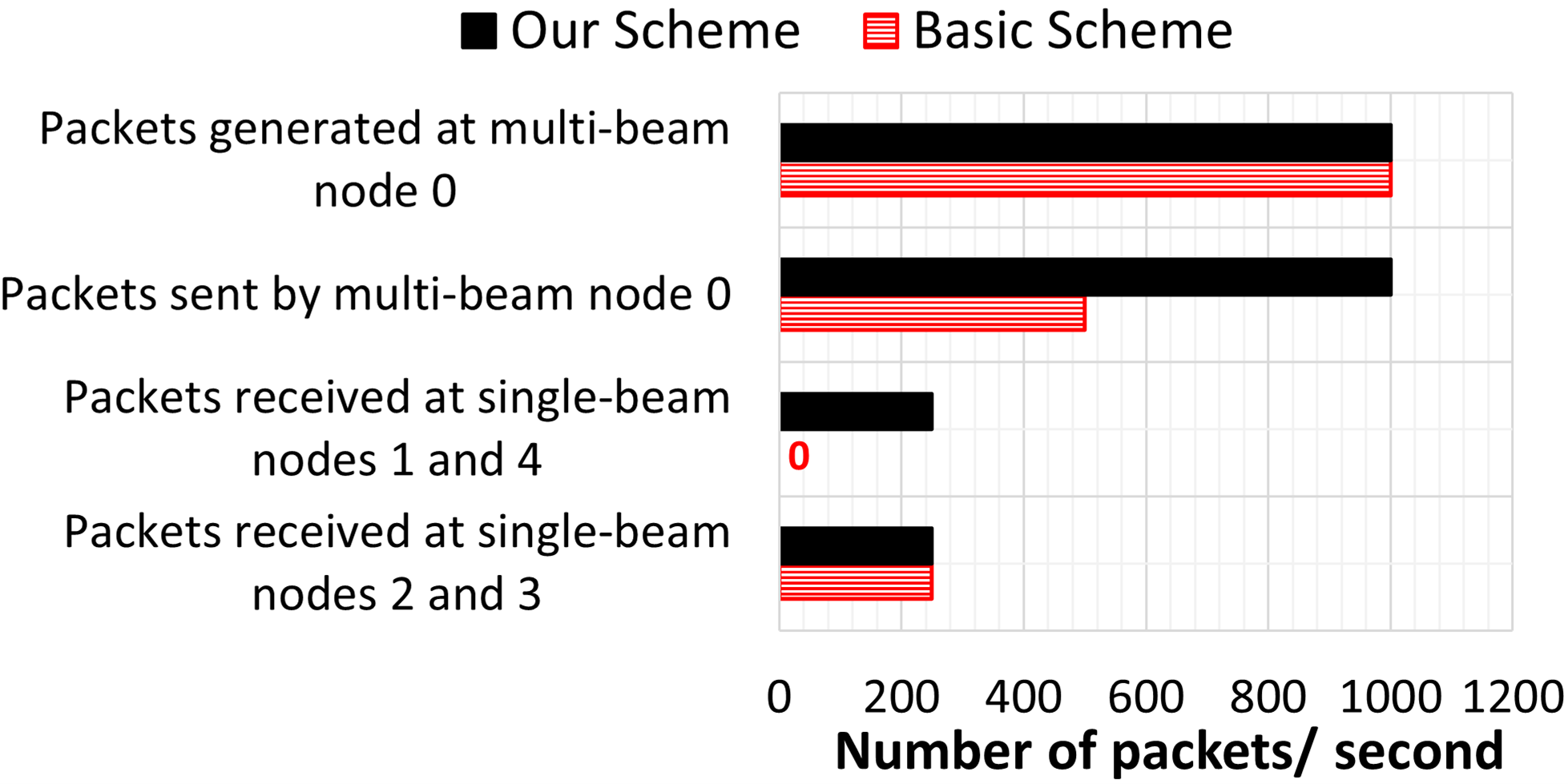}
    \caption{Data traffic at multi-beam transmitter node 0 and receiver nodes 1 through 4, in our proposed and basic multi-beam MAC schemes.}
    \label{Fig4}
\end{figure}

\subsubsection{Throughput Analysis when Transmitter is a Multi-beam Node}
\label{CPT}
Here, multi-beam node 0 transmits packets to nodes 1 through 4 concurrently on its four different beams (see Fig. \ref{Fig1} for network topology). The Euclidean distance of node 0 to nodes 2 and 3 is 2 km each, and node 0 to nodes 1 and 4 is 2.5 km each. Since nodes 1 to 4 are not equally distant from node 0, the difference in their propagation delay affects the synchronized data transmission and reception.

Multi-beam node 0 generates a total of 1000 packets per second (i.e., 250 packets per beam for nodes 1 through 4) as shown in the $1^{st}$ black bar from the top in Fig. \ref{Fig4}. Although the packets experience different propagation delays due to unequal distances between the transmitter node 0 and receiver nodes 1 through 4, the windowing mechanism helps the multi-beam node in achieving CPT. Each receiver node 1 through 4 receives 250 packets per second (see the $3^{rd}$ and $4^{th}$ black bars from the top in Fig. \ref{Fig4}). Therefore, the average flow throughput at each receiver node 1 through 4 is 3 Mbps in our proposed multi-beam MAC scheme in Table \ref{Table1}. 

\begin{table*}[ht]
\centering
\small
\caption{Performance Comparison of Both Schemes in Terms of Average Throughput}
\label{Table1}
\renewcommand{\arraystretch}{1.4}
\resizebox{0.85\textwidth}{!}{
\begin{tabular}{|c|c|c|c|}
\hline
\textbf{Multi-beam MAC Scheme $\downarrow$}                                      & \textbf{Single-Beam Nodes 1 and 4} & \textbf{Single-Beam Nodes 2 and 3} & \textbf{Multi-Beam Node 10} \\ \hline
\multicolumn{1}{|c|}{\textbf{Proposed Scheme}} & 3 Mbps                            & 3 Mbps                            & 12 Mbps                     \\ \hline
\multicolumn{1}{|c|}{\textbf{Basic Scheme}}    & 0                                 & 3 Mbps                            & 5.6 Mbps                    \\ \hline
\end{tabular}
}
\renewcommand{\arraystretch}{1}
\end{table*}

In the basic multi-beam MAC scheme, multi-beam node 0 sends a total of 1000 RTS packets concurrently on its four beams to nodes 1 through 4. However, it does not receive CTS from them at the same time due to its different distances from nodes 1 through 4. As a result, it receives a total of 500 CTS packets per second concurrently only from nodes 2 and 3. Therefore, it sends only 500 data packets per second (see $2^{nd}$ red bar from the top in Fig. \ref{Fig4}) out of the 1000 packets generated per second. Therefore, the average flow throughput at receiver node pairs 1, 4 and 2, 3 are 0 and 3 Mbps, respectively, in the basic multi-beam MAC scheme in Table \ref{Table1}. 

\begin{figure}[tbh]
    \centering
    \includegraphics[width=0.95\linewidth]{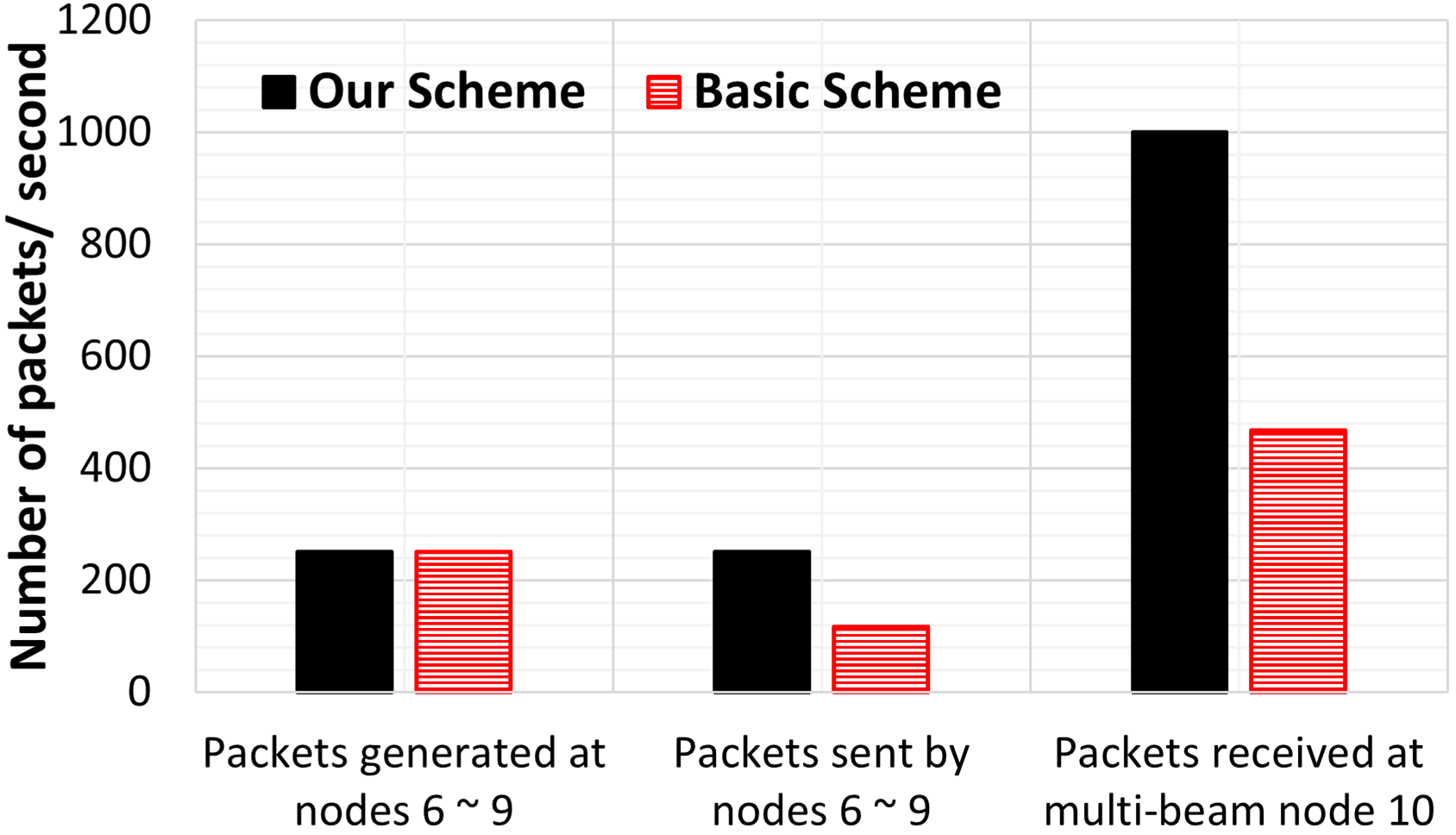}
    \caption{Data traffic at single-beam transmitter nodes 6 through 9, in our proposed and basic multi-beam MAC schemes.}
    \label{Fig5}
\end{figure}

\subsubsection{Throughput Analysis when Receiver is a Multi-beam Node}
\label{CPR}
Here, node 10 concurrently receives packets from nodes 6 through 9 on its four beams to test its CPR performance. The Euclidean distance of nodes 7 and 8 to node 10 is 2 km each, and nodes 6 and 9 to node 10 is 2.5 km each. Since nodes 6 through 9 are not equally distant from node 10, the difference in their propagation delay affects the synchronized data reception.

The single-beam transmitter nodes 6 through 9 generate 250 packets per second per node (see $1^{st}$ bars in Fig. \ref{Fig5}). In our proposed scheme, 250 packets per second are sent by each transmitter node (see $2^{nd}$ black bar). Here, all four transmitter nodes experience the same traffic behavior despite their different distances from the multi-beam receiver node 10 due to the use of windowing mechanism. Hence, the multi-beam node 10 receives a total of 1000 data packets per second (see $3^{rd}$ black bar). Therefore, the average flow throughput at the multi-beam receiver node 10 is 12 Mbps in our proposed multi-beam MAC scheme in Table \ref{Table1}.  

In the basic multi-beam MAC scheme, RTS packets from nodes 6 and 9, and 7 and 8 do not reach concurrently at the multi-beam receiver node 10 due to their unequal distances. As a result, nodes 6 and 9 (which are located farther from node 10 as compared to the nodes 7 and 8) double their backoff and retransmit packets towards node 10, which results in two locally synchronized transmitter node pairs (i.e., one pair of nodes 6 and 9, and another pair of nodes 7 and 8). Therefore, node 10 communicates with one node pair at a time while forcing the other node pair to wait by sending the N-CTS packet. Because of the alternate communication, each node transmits only 117 packets per second (see $2^{nd}$ red bar in Fig. \ref{Fig5}), and the remaining packets are dropped. Therefore, the average flow throughput at the receiver node 10 is 5.6 Mbps in the basic multi-beam MAC scheme in Table \ref{Table1}. 

\subsection{Performance Comparison for Mobile, Multihop Network Topology}
\label{MobileScenario}
We consider a 10 km x 10 km simulation area, where 50 airborne nodes fly under the Gauss-Markov mobility model \cite{Ref23} at 40 m/s. Each node has a 2 km transmission range. The number of flows considered are 1 and 3, where the source-destination pairs are randomly selected. The source and destination nodes use 4 antenna beams, whereas the remaining nodes use a single beam antenna. Since the source and destination nodes are connected via a multihop topology, the multipath-OLSR (optimized link state routing) protocol \cite{Ref24} is used to compute multiple node-disjoint routes, which are used by both schemes. Here, we assume that each node knows its 1-hop neighbors through periodic neighbor discovery. 

The performance is evaluated for three data rates (from 3 to 5 Mbps per flow) in terms of PDR, end-to-end delay and \% times our proposed scheme utilizes one or more extra routes as compared to the basic multi-beam MAC scheme. Here, the data rate represents the total data generated at the source node. The channel capacity is 3 Mbps. Each experiment is repeated 50 times and the window period is 9 µs. The remaining parameters are the same as discussed earlier.

\begin{figure}[b]
    \centering
    \includegraphics[width=0.8\linewidth]{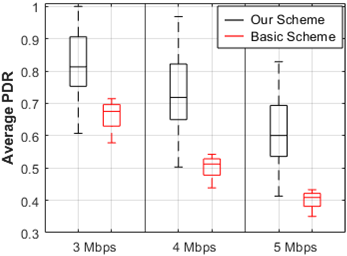}
    \caption{PDR comparison between our proposed and basic multi-beam MAC schemes for one source-destination pair.}
    \label{Fig6}
\end{figure}

\begin{figure}
    \centering
    \includegraphics[width=0.8\linewidth]{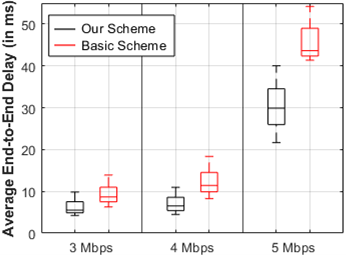}
    \caption{End-to-end delay comparison between our proposed and basic multi-beam MAC schemes for one source-destination pair.}
    \label{Fig7}
\end{figure}

\subsubsection{Performance Comparison for One Flow}
\label{OneFlow}
The PDR and end-to-end delay comparisons of our proposed and basic multi-beam MAC schemes are shown in Fig. \ref{Fig6} and \ref{Fig7}, respectively. Since the basic multi-beam MAC scheme cannot accommodate asynchronous packet arrival, it uses only one beam in the transmission or reception cycle, which increases the packets dropped. On the other hand, our proposed scheme accommodates the asynchronous packet arrival, which improves the channel access and resource management. This, in turn, improves its median PDR by up to 21\% as compared to the basic multi-beam MAC scheme. As the data rate increases, more packets are dropped due to buffer overflow, which decrease the PDR values in both schemes. However, our proposed scheme provides a significantly higher PDR at all the three data rates.

Unlike basic multi-beam MAC scheme, the source-destination pair in our proposed MAC scheme often utilizes more than one beam in the transmission and reception cycles and therefore achieves a lower end-to-end delay in Fig. \ref{Fig7}. The queuing delay increases with the data rate, which increases the end-to-end delay in both schemes. 

Note that only up to 1.5 Mbps flow data rate can be accommodated in a static 2-hop network topology with 3 Mbps channel capacity (assuming a zero control overhead and inter-flow interference) because the intermediate node requires half of the total time slots to forward the data packets on the second hop link. However, the maximum achievable flow throughput in a mobile, multi-hop network topology would be much lower than 1.5 Mbps. At a 5 Mbps data rate, a 4-beam source node generates 1.25 Mbps data per beam, which exceeds the maximum serviceable flow rate for the network setup considered in our simulation. It results in high congestion at the source node. Therefore, the average PDR decreases, and the end-to-end delay increases significantly at 5 Mbps data rate in Fig. \ref{Fig7}.

\begin{figure}[b]
    \centering
    \includegraphics[width=0.8\linewidth]{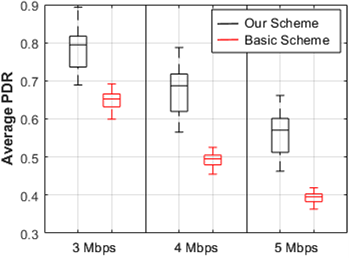}
    \caption{PDR comparison between our proposed and basic multi-beam MAC schemes for three source-destination pairs.}
    \label{Fig8}
\end{figure}

\begin{figure}
    \centering
    \includegraphics[width=0.8\linewidth]{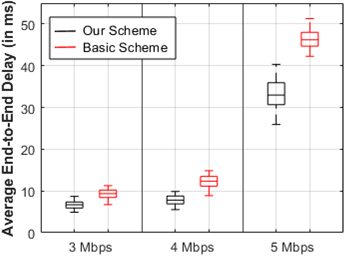}
    \caption{End-to-end delay comparison between our proposed and basic multi-beam MAC schemes for three source-destination pairs.}
    \label{Fig9}
\end{figure}

\begin{figure}[th]
    \centering
    \includegraphics[width=\linewidth]{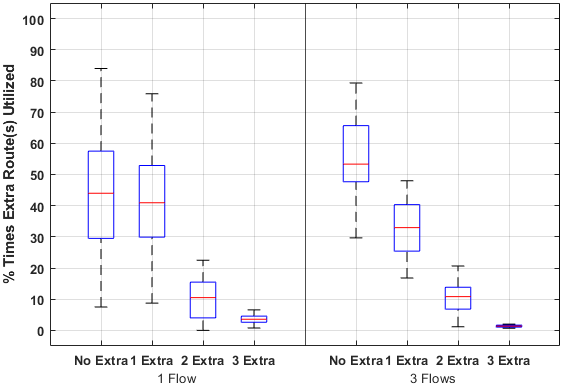}
    \caption{Percentage times when extra routes are utilized in our proposed multi-beam MAC scheme as compared to basic multi-beam MAC scheme with (\textit{i}) one flow (left) and (\textit{ii}) three flows (right).}
    \label{Fig10}
\end{figure}

\subsubsection{Performance Comparison for Multiple Flows}
\label{MultiFlows}
The PDR and end-to-end delay comparison of our proposed and basic multi-beam MAC schemes for 3 flows are shown in Fig. \ref{Fig8} and \ref{Fig9}, respectively. Due to the reasons discussed in the previous section, the average PDR is higher and the end-to-end delay is lower in our proposed scheme than the basic multi-beam MAC scheme at each data rate in Fig. \ref{Fig8}. 

Since the congestion in the network increases with the number of flows, the median PDR values at each data rate are lower for both schemes in Fig. \ref{Fig8} as compared to Fig. \ref{Fig6}. For the same reasons, the median end-to-end delay values in both schemes are higher in Fig. \ref{Fig9} than in Fig. \ref{Fig7}. 

\subsubsection{Improved Channel Utilization}
\label{ChannelUtilization}
Fig. \ref{Fig10} shows \% times when our proposed multi-beam MAC scheme utilizes extra route(s) as compared to the basic multi-beam MAC scheme. Note that both schemes have access to the same set of routes, but the basic multi-beam MAC scheme often utilizes only one route at a time due to asynchronous packet arrival. On the other hand, our scheme utilizes one or more extra routes up to 56\% times for the network with one flow (see Fig. \ref{Fig10} (left)). As the number of flows increases, the number of available node-disjoint routes decreases because an intermediate node (being a single beam node) can use only one beam at a time. Therefore, the \% times both schemes utilize the same number of routes increases with the number of flows (see Fig. \ref{Fig10} (right)). 

\section{CONCLUSION }
\label{Conclusion}
Compared to an omnidirectional or a single-beam directional antenna, a multi-beam node can achieve a throughput of up to \textit{m} times, by simultaneously communicating on its \textit{m} non-interfering beams. This requires a half-duplex multi-beam node to concurrently transmit or receive packets on its different beams. The existing multi-beam MAC schemes can facilitate concurrent communication when the transmitter nodes are locally synchronized. But they cannot accommodate asynchronous packet arrival at a multi-beam receiver node in a network where the nodes are randomly positioned and/or mobile, which increases the node deafness and MAC-layer capture problems, and thereby limits the flow throughput.

An asynchronous multi-beam MAC protocol for multi-hop wireless networks was presented in this paper. The following new features were introduced in the proposed scheme: (\textit{i}) A windowing mechanism to achieve the CPT and CPR when the packet arrival is asynchronous due to different packet sizes and varying propagation delays in a random and mobile topology, (\textit{ii}) A smart packet processing mechanism which reduces the node deafness, hidden terminals and MAC-layer capture problems, and (\textit{iii}) A new channel access mechanism which decreases the resource wastage and node starvation. The simulation results showed superior throughput, PDR, end-to-end delay and channel utilization performance of our proposed scheme as compared to the existing multi-beam MAC schemes for both static and dynamic network topologies.

\section{ACKNOWLEDGMENT OF SUPPORT AND DISCLAIMER}
The authors acknowledge the U.S. government’s support in the publication of this paper. This material is based upon the work funded by the U.S. AFRL (Air Force Research Lab), under Contract No. FA8750-14-1-0075 and FA8750-20-1-1005. Any opinions, findings and conclusions or recommendations expressed in this material are those of the author(s) and do not necessarily reflect the views of AFRL or the US government. 

\ifCLASSOPTIONcaptionsoff
    \newpage
\fi

\end{document}